# Printing Mosaics of Magnetically Programmed Liquid Crystal Directors for Reversibly Morphing Soft Matter

Yueping Wang[1†], Jongwon An[2†], Hongseok Kim[2], Sehui Jeong[2], Hyunggyu Kim[2], Jaesung Park[2], Seunggyu Ko[2], Jinho Son[2] and Howon Lee[2*]

[1]Department of Mechanical and Aerospace Engineering, Rutgers University, New Brunswick, NJ, USA

[2]Department of Mechanical Engineering, Institute of Advanced Machines and Design, Seoul National University, Seoul, Republic of Korea

[†]These authors contributed equally.

*Corresponding author (e-mail: howon.lee@snu.ac.kr)

**Abstract**

Liquid crystal elastomer (LCE) has been intensively utilized in 4D printing techniques to fabricate smart structures with reversible actuation on the basis of appropriate alignment of liquid crystal (LC) molecules. As a non-contact alignment strategy with a controllability of orientation, magnetic-field alignment has been rarely adapted in 4D printing of LCE because of its poor printing efficiency and demand on large field strength. Here, we report a digital light projection (DLP) system integrated with reorientable magnetic field to facilely print smart LCE structures. We propose a new LCE precursor solution that maintains a liquid crystalline nematic phase and an adequate flowability at room temperature. The resin prior to photopolymerization can be sufficiently aligned by a magnetic field with a strength of 500 mT in seconds without temperature elevating or cycling. Consequential printed structures are capable of presenting an impressive reversible thermal actuation of more than 30 %. The local and arbitrary magnetic-field alignment in layers during DLP printing is characterized, which renders us the ability to construct smart structures with more delicate LC alignments. Furthermore, we introduce the selective deformation of LCE structures with programmed molecular orientation using the photo-thermal effect. Our reported approach reveals the significant potential for fabricating morphing structures in various fields, including soft robotics, biomedical structures, and microelectronics.

**Keywords**: liquid crystal elastomers, DLP printing, magnetic alignment

4D printing, a rising additive manufacturing technique integrated by 3D printing with smart materials, has been attracting intense attention since its concept was brought forward years ago [1]. Conventionally, hydrogel [2-4] and shape memory polymer (SMP) [5-7] are two major smart materials employed in 4D printing to realize objects capable of transforming their shapes, properties, or functionalities in response to various external stimuli over time. Herein, hydrogel demonstrates a reversible shape-morphing but suffers from a low modulus confining its practical applications and a mandatory material transportation (e.g., water) through the network; SMP is solvent-free and owns a large stiffness tunability, but its actuation is irreversible. Liquid crystal elastomer (LCE), as an increasingly popular smart material recently, possesses the advantages beyond both hydrogel and SMP where a programmable and reversible deformation can be performed without material transportation or assistance of external loads, demonstrating great potentials in soft robotics [8-10], biomedical devices [11-13], and electronics [14].

The uniqueness of LCEs is owing to the primarily composed molecules: liquid crystal (LC, also called mesogen). LC is an intermediate state of matter that can exhibit both liquid fluidity and solid crystalline order and typically has a rod-like structure with in-series benzene-backbones, having shown many extraordinary characteristics such as flowability with long-range order, molecular cooperative motion, large birefringence, and alignment induced by exterior fields (e.g., mechanical, optical, magnetic, and electric) [15]. When introducing aligned LCs into elastomeric network, the developed LCE could demonstrate a mechanical [16], optical [17], and thermal [18] anisotropy as well as a large reversible actuation [19]. Particularly, the reversible actuation is mainly attributed to the phase transition in which LCs undergo a reorientation between a liquid crystalline order at the nematic state and a liquid disorder at the isotropic state.

The pivotal prerequisite for having a reversible actuation is to appropriately align LCs during the fabrication. So far, several strategies have been developed to reliably align LCs in LCEs, including surface alignment [20-24], mechanical stretch [9,19,25,26], mechanical shear [13,27-29], magnetic-field alignment [30-35], and electric-field alignment [36]. However, each strategy has more or less shortcomings that limit their extensive applications. For instance, surface alignment is merely confined to LCE films with a thickness of fewer than 100 μm so that bulk samples with sophisticated structures are impracticable; mechanical stretch is restricted from complex alignment fashions; mechanical shear is widely exploited in direct ink writing (DIW) printing in which the LC alignment is compulsorily coupled with the extruded filament path and the structural resolution is usually poor; magnetic-field alignment as a non-contact approach demands a long period of alignment according to previously reported studies, and temperature cycling is indispensable to LC alignment in some cases; electric-field alignment as another non-contact approach also requires a large field strength and lacks sufficient studies yet.

Diverse 4D printings of LCEs with programmable LC alignment have been reported by exploiting specific alignment method. DIW is the most widely studied 3D printing approach to fabricate LCE structures with complex local LC alignments. Basically, the LC alignments are induced by shear when the LCE filament is continuously extruded through a nozzle. Beyond the fundamental mechanism, many advancements of DIW of LCEs have been taken in recent years, such as realization of multi-LCEs printing [37], involvement of dynamic covalent bonds in DIW-printable LCEs [38], control of printing parameters to realize local graded actuations [39], integration of liquid metal with LCEs by purely mixing [40] or shell-core structure [41], and integration with other printing technologies [42,43]. Nevertheless, the intrinsic coupling of LC alignment along with the printing path confines a more fascinating actuation. And the printing procedure suffers from the extremely high viscosity of LCE ink, which possibly requires a high extrusion pressure and results in a depreciated printing resolution. Two-photon polymerization (2PP) utilizing surface alignment [44] or electric-field alignment [36] are also able to print LCE structures with fine LC alignments, but the printed structures are limited to micro size. Comparably, digital light projection (DLP) printing of LCEs can make sophisticated bulk structures as well as decoupling the printing direction with LC alignment orientations. A DLP system employed the shear of LCE resin generated by the continuous rotation of resin tray to align LCs in-situ, but the inevitable shear gradient in depth and the monotonous shear direction are still the limitations [45]. The ideal non-contact magnetic-field alignment has been applied to the DLP of LCEs in one study [31]. Although a conventional temperature cycling is avoided, an alignment dwell time of 5 min is still required, and the thermal actuation of printed structures is quite insignificant.

In this work, we propose a new LCE material formulation, and the prepared precursor solution is capable of maintaining a nematic state at room temperature and being facilely aligned under a sufficient magnetic field in seconds without temperature cycling or elevation. We build a customized DLP system with a reorientable magnetic field passing through the printing area. After characterizing material properties and the DLP printing, we are able to locally align LCs and lock the aligned network in printing to obtain smart structures with programmable and reversible thermal actuation. Correspondingly, several proof-of-concept demonstrations are presented with as-expected morphing upon heating and cooling. We also harness the photo-thermal effect to transfer the heat energy, enabling the selective control of shape morphing behavior. Our approach can be applied in various fields as it allows for the shape-morphing behavior of complex geometries.

**Results**

*DLP printing of magnetic-field aligned LCEs*

To fabricate LCE networks, we employed an acrylate-thiol photopolymerization reaction. In this process, 1,4-Bis-[4-(3-acryloyloxypropyloxy)benzoyloxy]-2-methylbenzene (RM257) and 2,2′-(Ethylenedioxy)diethanethiol (EDDET) were used as the mesogenic diacrylate monomer and the dithiol spacer, respectively (Fig. 1a). However, pure RM257 does not exhibit a nematic state unless the melting temperature ($T_m$) of 70 °C is exceeded. Therefore, we incorporated a non-reactive mesogenic solvent, 4-cyano-4-pentylbiphenyl (5CB) within RM257 powders at a 1:1 weight ratio. This allowed the mixture to maintain a nematic state at room temperature. Additionally, the nematic-isotropic transition temperature ($T_{NI}$) can also be suppressed from roughly 130 °C of the pure RM257 to lower than 85 °C of the RM257:5CB mixture. Subsequently, EDDET was mixed with the RM257:5CB mixture at a concentration of 60 mol% of the RM257. Phenylbis(2,4,6-trimethylbenzoyl) phosphine oxide as a photo-initiator (PI) and Sudan I as a photo-absorber (PA) were added at a concentration of 1 wt.% and 0.1 wt.% of the RM257 to control the photo-curable property. It's important to note that a thiol-acrylate mixture is known to have a limited shelf life due to the self-initiation of thiol-ene systems [46]. This unintended thiol-acrylate reaction was inhibited by including hydroquinone into the mixture at a concentration of 1 wt.% of the RM257. Early-stage material characterizations were conducted to display the Newtonian fluid feature of the mixture and its critical crystalline-to-nematic-transition temperature, denoted as $T_{CN}$, of 43 °C, while consumption of acrylate groups and thiol groups in the photopolymerization was also detected.

We built a customized DLP system with rotatable magnets to fabricate LCE structures with programmable LC alignments (Fig. 1b). Digital masks with a particular pattern were displayed on a digital micro-mirror device (DMD) that worked as a dynamic mask. Then UV light from a light emitting diode (LED) was reflected off the DMD and illuminated on the surface of the photocurable LCE solution in the glass cell through a projection lens. The pair of magnets with opposite poles facing each other can be rotated 360 ° in horizontal plane so that a controllable magnetic field orientation with a range of 180 ° was guaranteed in the printing area. Using the three-dimensional magnetic field model for a permanent magnet [47], the magnetic field distributions for various distances between the magnets were computed and confirmed the uniformity of the magnetic field strength in the UV curing area. Influenced by the magnetic field, free LC molecules in the precursor solution were reoriented and a following UV irradiation was able to initiate the photopolymerization to freeze the LC alignment in elastomeric network (Fig. 1c). Moreover, taking advantage of the selective projection capability of the DLP system, local programming of LC alignments was achievable in one layer (Fig. 1d). This was demonstrated by creating a bi-layer flower, where its first layer and the core of the second layer were cured without a magnetic field and all its six petals in the second layer exhibited radical LC alignments. The subsequent three pairs of petals were aligned by controlling the respective orientation of the magnetic field and then cured accordingly. This structure showed blooming behavior, with each petal undergoing bending actuation in response to temperature cycling.

The successful LC alignment in the fabricated LCE sample was verified visually via polarized optical microscopy (POM). A uniaxially aligned LCE film was observed between crossed polarizers (Fig. 1e). The captured area was dark when the LC alignment was parallel to either of the polarizers. When the crossed polarizers were rotated 45 °, however, a brighter area was obtained. This phenomenon is attributed to the well-known birefringence nature of LCs. We depicted the fundamental working principle of the reversible thermal deformation of proposed aligned LCEs in Fig. 1f: The as-fabricated LCE film owning a uniaxial LC alignment presented a nematic state at room temperature with all slender LC molecules approximately aiming to the same orientation. Upon heating above $T_{NI}$, the aligned film experienced a phase transition to an isotropic state, where the orientation order of LC molecules is destroyed. Macroscopically, the film performed a shrinkage in the LC alignment direction but an expansion in the directions perpendicular to the LC alignment. Then a cooling operation to room temperature recovered the initial shape of the film, demonstrating the reversibility of the thermal-induced deformation.

*Material properties and thermal actuation of magnetic-field aligned LCEs*

To better understand the magnetic-field aligned LCE, we systematically characterized mechanical and thermal properties of LCE samples with distinct LC alignments. In the fabrication, the parallel and perpendicular LC alignments were induced by intentionally allowing the magnetic field to pass through the LCE precursor solution contained in the glass cell in specific directions at room temperature, which were followed by UV illumination to lock the particular alignment orientation in LCE networks. Before performing temperature-responsive experiments, we gained the critical parameter, $T_{NI}$, of cured LCEs. Two distinct endothermic peaks were observed, one near 18 °C and another near 118 °C. The endothermic peak close to 18 °C corresponds to the transition temperature of 5CB molecules, which serve as porogens within the LCE structure. On the other hand, the endothermic peak near 118 °C marks the nematic to isotropic transition temperature of the elastomeric network.

When conducting iso-stress tests using a dynamic mechanical analyzer (DMA) upon the two types of aligned films, deformations toward opposite directions were observed along the longitudinal axis: the parallel aligned film showed a shrinkage over temperature (Fig. 2a), while the perpendicularly aligned film showed an expansion over temperature (Fig. 2b). The influence of magnetic field strength on the thermal-induced actuation strain was also surveyed. To illustrate, we fabricated both parallel and perpendicularly aligned LCE films under varying magnetic field strengths. All parallel aligned films behaved a negative actuation strain at 160 °C. With the increase of magnetic field strength from 100 mT to 500 mT, the actuation strain was improved from -14.1 % to -36.5 % monotonously. Similarly, all perpendicularly aligned films showed a positive actuation strain at 160 °C. The actuation strain was improved from 6.8 % to 29.8 % when increasing the magnetic field strength from 100 mT to 500 mT. Obviously, the orientation and strength of a magnetic field can both impose a straightforward effect on the thermal deformation of cured LCEs. Additionally, we assessed the thermal actuation stability of as-fabricated LCEs by monitoring the actuation strain of a parallel aligned film under a repeating temperature cycle in iso-stress test (Fig. 2c). The nearly stable strains of 0.2 % at room temperature (i.e., 20 °C) and 24.0 % at a high temperature (i.e., 120 °C) within a 5-cycle verified the thermal actuation stability of our proposed LCEs. It was noted that, while the viscous precursor solution was injected into the narrow glass cell, its spread induced a portion of LC molecules to be in-plane aligned. Thus, ultrasonication method was adopted when characterizing LCE properties.

To investigate the effectiveness of our magnetic-field alignment, iso-stress tests were performed to acquire the thermal-induced actuation strain in response to different waiting times in magnetic field before UV irradiation (Fig. 2d). All samples were made with parallel alignment at room temperature. Notably, a waiting time of 10 sec was sufficient to cause an actuation strain of -35.5 % at 160 °C. A further gradual increase of waiting time could barely lead to a significant improvement of actuation strain, which was evidenced by the actuation strain of -37.7 % at 160 °C for a waiting time of 10 min. Furthermore, by observing the color-changing outcome in the POM system with a 300 mT magnetic field strength, we could capture the birefringence variation of the LCE precursor solution. The color sequence in the one-second interval real-time POM results followed the Michel-Levy chart [48], which is a tool for identifying birefringence with a sample thickness and observed color. Overall, the results imply an instantaneous and adequate magnetic-field alignment of LCs is achievable at room temperature, so printing efficiency is dramatically improved.

*Characterization of DLP printing with magnetic field alignment*

It was essential to evaluate the built DLP printing system in order to explore the possibility of realizing LCE structures with a more delicate shape and LC alignment. We firstly conducted a curing depth study to find the relationship between the energy dosage and the curing depth of LCE precursor solution to confirm the printability. The growth of curing depth with the increase of energy dosage follows the stereolithography working curve equation (i.e., $C_d = D_p \ln(E/E_c)$), where $C_d$ is curing depth and $E$ is energy dosage. Two key constants, $D_p$ and $E_c$, are characteristic cure depth and critical energy required to initiate curing, respectively [49].

Carrying the evaluated printing parameters, we were able to print LCE structures with controlled dimensions and programmed LC alignment. For instance, bilayer LCE films with a non-alignment at the passive layer (150 μm) and a parallel alignment at the active layer (150 μm) were fabricated and showed predictable bending behaviors at rising temperatures (Fig. 3a). Their identical bending behaviors verified the agile LC alignment dependence on the magnetic field in the printing and implied the feasibility of endowing arbitrary programming of LC alignment into printed structures. Followingly, the in-layer local programming of LC alignment was investigated. We confirmed that multiple local programmings could be realized in one layer regardless of their sequence. It makes printing LCE structures with sophisticated thermal actuation possible.

Exploiting the versatile programming of LC alignment and selective photopolymerization in the DLP printing, we came up with several bilayer morphing structures that demonstrated interesting thermal-induced deformations. It should be noted that all bilayer structures in this section possess a total thickness of 200 μm and a reversible thermal-induced morphing. The strategy involving the same location but various magnetic field orientations was demonstrated (Fig. 3b). Three bilayer films were printed with their active layer aligned by different magnetic field orientations (0 °, 30 °, and 90 ° with respect to the film length direction). Similarly, through the heating/cooling cycle, we clearly observed distinct reversible bending actuations that corresponded to the respective LC alignments of the three films. The validation of distinct magnetic field alignment for each layer and each location within a layer was further confirmed using two slender LCE films (Fig. 3c). We applied the magnetic-field alignment to the entirely one layer of the left film and to different halves of both layers in the right film. Upon heating to 120 °C in an oven, the left film exhibited uniform upward bending, while the right film displayed an "S"-shaped deformation due to the opposite bendings on its two sides. Likewise, we performed finite-element analysis (FEA) simulations on these structures, which output the deformations in good agreement with their experimental counterparts.

The mimicry of flower blooming is a popular demonstrating approach, and there have been many studies presenting versatile blooming-like behaviors [39,50]. Since we had validated the in-layer arbitrary LC alignments induced by the magnetic field, it was facile to utilize this advantage to manufacture bilayer flower structures with distinct blooming behaviors. The flower structures were designed with six petals and their designated LC alignments (Fig. 3d). Specifically, one layer was set as a non-aligned passive layer that experienced an ignorable deformation upon heating. In another layer, however, the first design visualized the LC alignment following the respective midvein direction of petals. On the basis of the first design, we rotated all LC alignments by +20 ° (counterclockwise) with respect to their midveins in the second design. Furthermore, the corresponding rotations of LC alignments by +20 ° or -20 ° in the third and fourth designs were attempted. All petals curled upward upon heating to 120 °C and twisted in their deflecting LC alignment directions. As a result, distinct blooming demonstrations were realized with ease, and their morphing reversibility could be observed. In addition, FEA simulations on these four flower structures displayed good alignment with the experimental demonstrations.

*Diverse shape-morphing of magnetic-field aligned LCE structures*

LCE structures were previously demonstrated to actuate the deformation of kirigami metastructures or 2D-to-3D mesostructures [14,25,28]. However, their aligned LCEs serve as exterior actuators to be assembled with other reconfigurable parts, which demands more labors. Taking advantage of the rapid prototype process and the arbitrary magnetic-field LC alignment of our built system, it is promising to directly impose the reversible LCE actuation into the advanced structures.We demonstrated several proof-of-concept prototypes (Fig. 4). To begin with, the kirigami sample was easily printed with a thickness of 300 μm and a monotonous alignment (Fig. 4a). When fixing its boundaries, the thermal-induced shrinkage along the length was confined so that a kirigami transformation occurred as expected. A similar FEA result was obtained when applying the same boundary conditions.

Moreover, we constructed two buckling structures to demonstrate the 2D-to-3D deformation (Fig. 4b, c). To avert a simultaneous bending of the actuating frame when enabling the central buckling, we thickened the surrounding frame to improve its bending rigidness. Frames were aligned unidirectionally in the 1D buckling structure and bidirectionally in the 2D buckling structure to implement the buckling mechanis,. When fabricating the frame structures, a 300 µm thickness spacer was used to print one layer, and three layers were stacked to achieve a total frame thickness of 900 µm. The inner buckling part was printed at a thickness of 200 um, which was achieved in the first layer with a curing time of 4 sec. The printed structures exhibited buckling actuation at 160 °C, resulting from the contraction deformation of aligned frames. FEA simulations were conducted to validate the shape of the deformed buckling structures. The results showed similar deformed shape between the actual configurations and the simulations.

One more exemplar fabricated was an octagonal crown with inner strands and a ring structure (Fig. 4d). Each side of the octagonal frame was aligned at 0 °, 45 °, 90 °, and 135 °. Printed thicknesses were similar to those for the aforementioned 1D and 2D buckling structures. Upon heating to 160 °C, the aligned octagonal frame underwent contraction, leading to an upward bending deformation of the internal strands. This deformation of strands was the result of maintaining the shape of the ring structure. Subsequent FEA simulation yielded the same result for the crown structure.

The final demonstration was a flying fish-like shape-morphing structure (Fig. 4e). Unlike the previous examples, this flying fish-like structure introduced a mechanism where the center bar contracts, resulting in the deformation of the outer wing structure. In this structure, the center bar was fabricated with magnetic field alignment by stacking three layers, each with a thickness of 300 µm. The wing-like structure attached to the bar was printed at a thickness of 300 µm without a magnetic field. Upon heating to 160 °C, the center bar structure underwent thermal-induced contraction, leading to the out-of-plane deformation of the wings attached to the bar. This upward deformation occurs because the moment of inertia for out-of-plane deformation is higher than that for in-plane deformation. The FEA simulation result also exhibited good agreement with the actual deformed geometry.

*Photo-thermal-activated LCE structures with magnetic field alignment*

In addition to transporting direct heat energy to randomize aligned molecules in LCE structures, another promising approach takes advantage of the photo-thermal effect, in which light energy is absorbed by a colored target and converted to heat to induce expected functionalities. This method was frequently employed in the literature to realize facile manipulation with complex deformation of structures, either by doping photo-thermal agents into sample networks [4,13,29,51] or by introducing colored skins on sample surfaces [52,53]. Inspired by the work of Liu et al. [53], we endowed LCE structures with the feature of selective photo-thermal actuation by simply painting arbitrary areas of the structure with different colors, by doing so we averted bringing photo-thermal agents into LCE network. Because PA in the resin was able to result in an inherently yellow color of printed LCEs, which was not wanted, we eliminated PA when fabricating LCEs in this section so that printed structures possessed a translucent and near-white color and insignificant light absorption.

Beneficial from the capability of local LC programming and selective photo-curing of our DLP system, we designed a soft crawler composed of three LCE layers (Fig. 5a). Particularly, the 150-μm bottom layer had designated LC alignments of 20 ° and -20 ° about the central axis on two limb areas. During actuation, the 150-μm non-aligned middle layer enabled the downward bending of the two limbs and the 600-μm non-aligned top layer prevented unwanted deformation of the body of the crawler. The two limbs of the prototype were painted with black color to eable the photo-activation. A blue LED was used to induce the photo-thermal actuation of the crawler. After turning the light on, the thermal video proved a swift and adequate temperature elevating (a change from ~25 °C at 0 sec to ~115 °C at 20 sec) and a consequently significant bending deformation of the two limbs within 20 sec. In the crawling demonstration, we applied a periodic actuation with 20-sec light-on and 30-sec light-off and analyzed the motion mechanism in one period (Fig. 5b). Initially, the crawler was still on a rugged surface (Fig. 5b(i)). Switching the light on induced the bending of limbs (Fig. 5b(ii)). Because the friction of the limb tip was larger than that of the tail, the limb tip as the support point enabled the forward move of the entire body (Fig. 5b(iii)). A following switching the light off released the bending of limbs (Fig. 5b(iv)). However, the friction of the tail prevailed at this moment, so a further forward move of the crawler was guaranteed (Fig. 5b(v)). By repeating the aforementioned process, we could achieve the continuous forward crawling of our LCE crawlers. The displacement of the head of the crawler was extracted from the actuation video with respect to time, which indicated a total displacement of 13.7 mm after 11 periods (i.e., 550 sec). Nevertheless the movement of our crawler is not very efficient, it reveals the success of the photo-thermal actuation of our proposed LCE structures. And a replacement with other more effective LED lights could certainly improve the crawling performance.

Direct heating causing simultaneous deformations everywhere on an LCE structure is not always desired, especially when local actuations are needed in some scenarios. Therefore, we investigated the selective photo-thermal actuation of LCEs by preparing single-layer non-aligned LCE films with a thickness of 300 μm that were painted with different colors: red, yellow, green, cyan, blue, and pink. Under either red, green, or blue LEDs, the surface temperature on all films after a 60-sec illumination was monitored (Fig. 5c). From the radar plot, it was explicit that the blue light overwhelmed others to heat the yellow-painted film up to 131 °C and the cyan-painted film could be exclusively heated up to 105 °C under the red light. Due to the measured low light intensity of the green light compared to others, its effectiveness for heating colored samples was diminished. Nevertheless, we could observe its heating performance on the pink-painted film, a lifting temperature of 87 °C was witnessed under the green light.

Based on the collected information, we demonstrated selective photo-thermal actuations using a bilayer flower with three leaves (Fig. 5d). Each leaf has a midvein-alignment of LCs on the top layer, and they were painted with pink, yellow, and cyan color, respectively. Depending on the type of applied LED light, the corresponding unique leaf would be actuated and curl upwards. Additionally, by utilizing the selective photo-thermal actuation, the two-way light activated LCE crawler was exhibited (Fig. 5e). The molecular aligned two sets of limbs were painted with cyan and yellow ink, respectively. When the red light was irradiated, only cyan-painted limbs were activated and traveled to the right. 40 seconds of light irradiation and 30 seconds of cooling were employed in the procedure. On the other hand, a blue light was irradiated for 20 seconds and then cooled for 30 seconds to activate the yellow-painted limbs and permit leftward mobility. The two-way locomotion of the crawler on a rough surface was observed depending on the color of the irradiated light. Consequently, we demonstrated the decoupling of complex geometry creation, unconstrained molecular orientation programming, and selective photo-thermal actuation.

**Conclusion**

In this study, we focus on the amelioration of magnetic-field alignment of LCE in 3D printing. A new material formulation composed of acrylate-based RM257, thiol-based EDDET, and non-reactive 5CB is proposed. The prepared LCE precursor solution can maintain a nematic state and possess a suitable viscosity at room temperature so that we are capable of applying a magnetic-field alignment on the LCE resin at the ambient temperature. We unravel the mechanical anisotropy with respect to the alignment direction, a thermal stability at a rising temperature (e.g., 160 °C), a rapid (as short as 10 sec) and sufficient LC alignment, and an actuation repeatability of cured LCEs through tests. Other systematic investigations about the impacts of 5CB and EDDET on sample performances are performed and revealed the necessity of possessing proper concentrations of them.

Our customized DLP printing system is integrated with a pair of magnets to enable the arbitrary LC alignment (180 ° in-plane) along with the selective photopolymerization. The local LC alignment has been proved to agilely depend on the magnetic field direction and be independent on the curing sequence, allowing for versatile engraving of LC alignment patterns in layers. We demonstrate several proof-of-concept LCE prototypes. Along with the FEA simulations, the thermal-induced behavior of our proposed examples is displayed and able to comply with our expectations. We also explore the possibility of simple photo-thermal actuation of LCE structures by painting colored inks on designated structural surfaces. This approach enables the effective untethered motion of color-painted structures, such as the flower structure and soft crawlers.

However, the demonstrated samples in our work are still structurally simple because the number of layers is limited in the current 3D printing. Some future efforts should be taken on making more sophisticated structures by further upgrading the DLP system. Additionally, the magnetic field orientation is confined in horizontal plane with only one degree of freedom (DOF). By adding more DOFs in the future upgrade, LCE structures with 3D elegant LC alignment can be expected. LCE material formulation should also be studied further to offer a more controllable, magnetic-field-alignable, and stable precursor solution adaptive to the DLP printing. Nevertheless, our developed LCE resin with a capability of being fast magnetic-field-aligned in an ambient environment provides significant insights into the 3D printing of LCEs as well as paving the way to more potential applications in diverse fields including soft robotics, biomedical devices, microelectronics, and deployable systems.

**Methods**

*Materials*

The liquid crystal elastomer (LCE) precursor solution consisted of 1,4-Bis-[4-(3-acryloyloxypropyloxy)benzoyloxy]-2-methylbenzene (RM257, CAS 174063-87-7, Daken Chemical Limited) as a diacrylate monomer, 4-cyano-4-pentylbiphenyl (5CB, CAS 40817-08-01, Daken Chemical Limited) as a non-reactive mesogenic solvent, 2,2′-(Ethylenedioxy)diethanethiol (EDDET, Product No. 465178, Sigma-Aldrich, Inc.) as a dithiol spacer, phenylbis(2,4,6-trimethylbenzoyl) phosphine oxide (Product No. 511447, Sigma-Aldrich, Inc.) as a photo-initiator (PI), Sudan I (Product No. 103624, Sigma-Aldrich, Inc.) as a photo-absorber (PA), and Hydroquinone (Product No. H9003, Sigma-Aldrich, Inc.) as a polymerization inhibitor. Trichloro(1H,1H,2H,2H-perfluorooctyl)silane (Product No. 448931, Sigma-Aldrich, Inc.) was used to make a low surface energy substrate. All materials were used as received without further purification.

*Formulation of the precursor solution*

The preparation was completed in an amber glass vial. 2 g of RM257 powder and 0.02g of Hydroquinone (concentration of 1 wt.% of the RM257) was dissolved in 2 g of 5CB solution (same amount as RM257) by stirring at 90 °C for 60 min. 0.02 g of PI and 0.002 g of PA were added at a concentration of 1 wt.% and 0.1 wt.% of the RM257 to the solution. After cooling down to room temperature, 0.3717 g of EDDET was mixed with the solution (at a concentration of 60 mol% of the RM257) and stirred at room temperature for 10 min. Then the precursor solution was ready to use. Additionally, 0.1858 g and 0.5575 g of EDDET was mixed with the solution when a concentration of 30 mol% and 90 mol%, respectively, of the RM257, was required.

*DLP printing systems with magnetic-field assistance*

A customized DLP system was built with the following major components: a UV LED (365 nm, L10561, Hamamatsu), a projection lens (105 mm, f/4, UV-micro-apo, 114019, Coastal Opt.), a digital micro-mirror device (DMD) (DLPLCR6500EVM, Texas Instruments), a motorized linear stage (LTS150, Thorlabs), and a 3D printed sample holder. Two permanent neodymium magnets (NB064-N52, Applied Magnets) were mounted on a custom-made rotatable stage to enable the orientation control of the magnetic field in the printing space. A custom-written LabVIEW (National Instruments) script was used to control the printing work. The entire DLP system was kept in a UV blocking enclosure.

Prior to a printing job with the customized DLP system, the sample holder was lifted. A glass cell with a particular gap was washed with ethanol and cleaned to dry. The LCE precursor solution was injected into the glass cell, followed by laying the glass cell into the sample holder. Next, the sample holder was lowered down to the projection plane. After adjusting the magnetic field orientation (the magnetic field strength was kept 500 mT unless otherwise specified) and giving a sufficient waiting time (30 sec unless otherwise specified), patterned UV light was projected on the glass cell to photo-polymerize the precursor solution. If a second layer was required, the glass cell was detached to double the gap's thickness and refill new precursor solution while maintaining the previously printed layer in position. Then the projection process was repeated. Otherwise, the glass cell was removed from the lifted sample holder, followed by washing away the residual solution with deionized (DI) water. After DLP printing, samples were dried in the air until surface water evaporated. Subsequently, the samples were post-cured in a UV oven (365 nm, CL-1000L, UVP) for 10 min for each side.

To progress a printing job with the second customized DLP system, a glass slide was washed with ethanol and dry air. After then, the glass slide was coated with Trichloro(1H,1H,2H,2H-perfluorooctyl)silane in the desiccator under vacuum for 30 min. The LCE precursor solution was poured onto the coated glass slide with spacers of a specific thickness and then covered with another coated glass slide (to detach cured LCE from the glass slide) or non-coated glass slide (to fix the cured LCE first layer on the glass slide for bi-layer LCE samples). Next, the glass cell was put into the sample holder tuned to projection plane. After adjusting the magnetic field and waiting for a while, UV image was projected on the glass cell. When a second layer was required, new precursor solution was poured onto the non-coated glass slide with cured LCE first layer and spacers of total layer thickness and then covered with another coated glass slide. Next, the molecular orientation programming and UV image irradiation were conducted. After finishing the projection process, the cured LCE structures was removed from the glass cell and washed away the residual solution with deionized (DI) water. The samples were dried in the air to remove the water on the surface and post-cured in a UV oven (365nm, CL-3000L, UVP) for 10 min for each side.

*Demonstration of photo-activated crawler*

The limbs of the as-printed crawler were painted with black color using an oil-based marker (Black, Sharpie). The crawler was ready to demonstrate after the ink was dried out. 50-W blue LED (460-470 nm, Odlamp) was used as the heating source and placed vertically away from the surface of the crawler by 20 mm in the demonstration. A thermal camera (TIM450, Micro-Epsilon) was placed out of the light irradiation area to monitor the temperature evolution of the limbs of the crawler in the light heating phase. A MATLAB (R2019a, MathWorks) code was compiled to extract the locomotion of the crawler under a periodic light actuation. In the code, the captured video was imported, and frames were analyzed to record the displacement of the head of the crawler over time.

*Temperature investigation of painted LCEs under different LED lights*

Six single-layer non-aligned LCE films with a thickness of 300 μm were prepared and painted with red, yellow, green, cyan, blue, and pink colors, respectively, using various oil-based markers (Red, yellow, cyan, and blue, Sharpie; Green and pink, Uni-Ball). 50-W red (620-625 nm), 50-W green (520-525 nm), 50-W blue (460-470 nm) LEDs (purchased from Odlamp) were used as the heating sources. The 50-W red and 50-W green LEDs were placed away from the surface of the films by 10 mm and the 50-W blue LED was placed away from the surface of the films by 20 mm to provide reasonable high light intensity and retain a safe distance for demonstration. The energy intensity measured using an optical power meter (FieldMaxII-TO, Coherent) from the corresponding setting distance of red, green, and blue LEDs is 522.9 mW/cm$^2$, 323.0 mW/cm$^2$, and 479.8 mW/cm$^2$. A thermal camera (FLIR C2, FLIR® Systems, Inc.) was used to measure the temperature of films at a steady state under different LED lights.

**Acknowledgments**

The authors acknowledge the financial support from the National Research Foundation of Korea (NRF) Grants funded by the Korean Government (MSIT) (RS-2023-00208052 and No. 1711154190) and from the SNU Creative-Pioneering Researchers Program.

**Competing interests**

The authors declare no conflict of interest.

**References**

1    Tibbits, S. 4d Printing: Multi-Material Shape Change. *Archit Design* **84**, 116-121 (2014).

2    Han, D., Lu, Z. C., Chester, S. A. & Lee, H. Micro 3D Printing of a Temperature-Responsive Hydrogel Using Projection Micro-Stereolithography. *Sci Rep* **8** (2018).

3    Han, D. *et al.* Soft Robotic Manipulation and Locomotion with a 3D Printed Electroactive Hydrogel. *Acs Appl Mater Inter* **10**, 17512-17518 (2018).

4    Han, D., Wang, Y. P., Yang, C. & Lee, H. Multimaterial Printing for Cephalopod-Inspired Light-Responsive Artificial Chromatophores. *Acs Appl Mater Inter* **13**, 12735-12745 (2021).

5    Ge, Q. *et al.* Multimaterial 4D Printing with Tailorable Shape Memory Polymers. *Sci Rep* **6** (2016).

6    Yang, C. *et al.* 4D printing reconfigurable, deployable and mechanically tunable metamaterials. *Mater Horiz* **6**, 1244-1250 (2019).

7    Yang, C. *et al.* 4D-Printed Transformable Tube Array for High-Throughput 3D Cell Culture and Histology. *Adv Mater* **32** (2020).

8    He, Q. G. *et al.* Electrically controlled liquid crystal elastomer-based soft tubular actuator with multimodal actuation. *Sci Adv* **5** (2019).

9    He, Q. G. *et al.* Electrospun liquid crystal elastomer microfiber actuator. *Sci Robot* **6** (2021).

10	Kim, D. S., Lee, Y. J., Kim, Y. B., Wang, Y. C. & Yang, S. Autonomous, untethered gait-like synchronization of lobed loops made from liquid crystal elastomer fibers via spontaneous snap-through. *Sci Adv* **9** (2023).

11	Wu, J. *et al.* Liquid Crystal Elastomer Metamaterials with Giant Biaxial Thermal Shrinkage for Enhancing Skin Regeneration. *Adv Mater* **33** (2021).

12	Wang, Y. P., Liao, W., Sun, J. H., Nandi, R. & Yang, Z. Q. Bioinspired Construction of Artificial Cardiac Muscles Based on Liquid Crystal Elastomer Fibers. *Adv Mater Technol* **7** (2022).

13	Tasmim, S. *et al.* Liquid crystal elastomer based dynamic device for urethral support: Potential treatment for stress urinary incontinence. *Biomaterials* **292** (2023).

14	Li, Y., Luo, C. Q., Yu, K. & Wang, X. J. Remotely Controlled, Reversible, On-Demand Assembly and Reconfiguration of 3D Mesostructures via Liquid Crystal Elastomer Platforms. *Acs Appl Mater Inter* **13**, 8929-8939 (2021).

15	Demus, D., Goodby, J., Gray, G. W., Spiess, H.-W. & Vill, V. *Handbook of liquid crystals set*.  (Wiley Online Library, 1998).

16	Merkel, D. R., Traugutt, N. A., Visvanathan, R., Yakacki, C. M. & Frick, C. P. Thermomechanical properties of monodomain nematic main-chain liquid crystal elastomers. *Soft Matter* **14**, 6024-6036 (2018).

17	Stapert, H. R., del Valle, S., Verstegen, E. J. K., van der Zande, B. M. I. & Stallinga, S. Photoreplicated anisotropic liquid-crystalline lenses for aberration control and dual-layer readout of optical discs. *Adv Funct Mater* **13**, 732-738 (2003).

18	Shin, J. *et al.* Thermally Functional Liquid Crystal Networks by Magnetic Field Driven Molecular Orientation. *Acs Macro Lett* **5**, 955-960 (2016).

19	Yakacki, C. M. *et al.* Tailorable and programmable liquid-crystalline elastomers using a two-stage thiol-acrylate reaction. *Rsc Adv* **5**, 18997-19001 (2015).

20	Guin, T. *et al.* Layered liquid crystal elastomer actuators. *Nat Commun* **9**, 2531 (2018).

21	Fowler, H. E., Rothemund, P., Keplinger, C. & White, T. J. Liquid Crystal Elastomers with Enhanced Directional Actuation to Electric Fields. *Adv Mater* **33** (2021).

22	Hebner, T. S. *et al.* Discontinuous Metric Programming in Liquid Crystalline Elastomers. *Acs Appl Mater Inter* **15**, 11092-11098 (2023).

23	Hebner, T. S., Korner, K., Bowman, C. N., Bhattacharya, K. & White, T. J. Leaping liquid crystal elastomers. *Sci Adv* **9** (2023).

24	McCracken, J. M., Hoang, J. D., Herman, J. A., Lynch, K. M. & White, T. J. Millimeter-Thick Liquid Crystalline Elastomer Actuators Prepared by-Surface-Enforced Alignment. *Adv Mater Technol* **8** (2023).

25	Zhang, M. C. *et al.* Liquid-Crystal-Elastomer-Actuated Reconfigurable Microscale Kirigami Metastructures. *Adv Mater* **33** (2021).

26    Lewis, K. L. *et al.* Programming Orientation in Liquid Crystalline Elastomers Prepared with Intra-Mesogenic Supramolecular Bonds. *Acs Appl Mater Inter* **15**, 3467-3475 (2023).

27    Lu, X. L. *et al.* 4D-Printing of Photoswitchable Actuators. *Angew Chem Int Edit* **60**, 5536-5543 (2021).

28    Kotikian, A., Truby, R. L., Boley, J. W., White, T. J. & Lewis, J. A. 3D Printing of Liquid Crystal Elastomeric Actuators with Spatially Programed Nematic Order. *Adv Mater* **30** (2018).

29    Wang, Y. C. *et al.* 3D-Printed Photoresponsive Liquid Crystal Elastomer Composites for Free-Form Actuation. *Adv Funct Mater* **33** (2023).

30    Cui, J. X. *et al.* Bioinspired Actuated Adhesive Patterns of Liquid Crystalline Elastomers. *Adv Mater* **24**, 4601-4604 (2012).

31    Tabrizi, M., Ware, T. H. & Shankar, M. R. Voxelated Molecular Patterning in Three-Dimensional Freeforms. *Acs Appl Mater Inter* **11**, 28236-28245 (2019).

32    Li, S. C. *et al.* Controlling Liquid Crystal Orientations for Programmable Anisotropic Transformations in Cellular Microstructures. *Adv Mater* **33** (2021).

33    Yao, Y. X. *et al.* Multiresponsive polymeric microstructures with encoded predetermined and self-regulated deformability. *P Natl Acad Sci USA* **115**, 12950-12955 (2018).

34    Li, S. C. *et al.* Self-regulated non-reciprocal motions in single-material microstructures. *Nature* **605**, 76-83 (2022).

35  Li, S., Aizenberg, M., Lerch, M. M. & Aizenberg, J. Programming Deformations of 3D Microstructures: Opportunities Enabled by Magnetic Alignment of Liquid Crystalline Elastomers. *Acc. Mater. Res.* (2023).

36  Munchinger, A. *et al.* Multi-Photon 4D Printing of Complex Liquid Crystalline Microstructures by In Situ Alignment Using Electric Fields. *Adv Mater Technol* **7** (2022).

37  Kotikian, A. *et al.* Untethered soft robotic matter with passive control of shape morphing and propulsion. *Sci Robot* **4** (2019).

38  Davidson, E. C., Kotikian, A., Li, S. C., Aizenberg, J. & Lewis, J. A. 3D Printable and Reconfigurable Liquid Crystal Elastomers with Light-Induced Shape Memory via Dynamic Bond Exchange. *Adv Mater* **32** (2020).

39  Wang, Z. J. *et al.* Three-dimensional printing of functionally graded liquid crystal elastomer. *Sci Adv* **6** (2020).

40  Ambulo, C. P., Ford, M. J., Searles, K., Majidi, C. & Ware, T. H. 4D-Printable Liquid Metal-Liquid Crystal Elastomer Composites. *Acs Appl Mater Inter* **13**, 12805-12813 (2021).

41  Kotikian, A. *et al.* Innervated, Self-Sensing Liquid Crystal Elastomer Actuators with Closed Loop Control. *Adv Mater* **33** (2021).

42  Peng, X. R. *et al.* Integrating digital light processing with direct ink writing for hybrid 3D printing of functional structures and devices. *Addit Manuf* **40** (2021).

43      Peng, X. R. *et al.* 4D Printing of Freestanding Liquid Crystal Elastomers via Hybrid Additive Manufacturing. *Adv Mater* **34** (2022).

44      Guo, Y. B., Shahsavan, H. & Sitti, M. 3D Microstructures of Liquid Crystal Networks with Programmed Voxelated Director Fields. *Adv Mater* **32** (2020).

45      Li, S. *et al.* Digital light processing of liquid crystal elastomers for self-sensing artificial muscles. *Sci Adv* **7** (2021).

46      Hoyle, C. E., Lee, T. Y. & Roper, T. Thiol-enes: Chemistry of the past with promise for the future. *J Polym Sci Pol Chem* **42**, 5301-5338 (2004).

47      Weir, G., Chisholm, G. & Leveneur, J. The Magnetic Field About a Three-Dimensional Block Neodymium Magnet. *Anziam J* **62**, 386-405 (2020).

48      Sorensen, B. E. A revised Michel-Levy interference colour chart based on first-principles calculations. *Eur J Mineral* **25**, 5-10 (2013).

49      Gibson, I., Rosen, D. & Stucker, B. *Additive Manufacturing Technologies*. 2 edn, (Springer, 2015).

50      Gladman, A. S., Matsumoto, E. A., Nuzzo, R. G., Mahadevan, L. & Lewis, J. A. Biomimetic 4D printing. *Nat Mater* **15**, 413-418 (2016).

51      Li, Z. *et al.* Polydopamine nanoparticles doped in liquid crystal elastomers for producing dynamic 3D structures. *J Mater Chem A* **5**, 6740-6746 (2017).

52      Tian, H. M. *et al.* Polydopamine-Coated Main-Chain Liquid Crystal Elastomer as Optically Driven Artificial Muscle. *Acs Appl Mater Inter* **10**, 8307-8316 (2018).

53      Liu, Y., Shaw, B., Dickey, M. D. & Genzer, J. Sequential self-folding of polymer sheets. *Sci Adv* **3** (2017).

**Fig. 1. Material composition and DLP printing of LCEs with magnetic-field alignment. a**, Chemical structures of mesogenic diacrylate monomer RM257, dithiol spacer EDDET, and non-reactive mesogenic solvent 5CB in the formulated LCE precursor solution. **b**, Schematic

illustration of the DLP printing system with controllable magnetic field. The magnetic field is generated by two permanent neodymium magnets with opposite poles facing one another, which is able to horizontally pass through the printing area with a rotation controllability of 360°. **c**, Schematic illustration of the magnetic-field alignment of LCE precursor solution. LCs are aligned to roughly one orientation by following the magnetic field line, followed by UV-cured polymeric network to freeze the alignment orientation. **d**, Schematic illustration of the in-plane arbitrary magnetic-field alignment. Taking advantage of the rotatable magnetic field and the selective UV photopolymerization, LC alignment toward distinct orientations can be achieved. **e**, POM images of uniaxially aligned LCE showing the successful LC alignment in the network. **f**, Reversible thermal deformation of uniaxially aligned LCE film. At the nematic state ($T < T_{NI}$), LCs point to the approximately same orientation in the film. Upon heating above $T_{NI}$, at the isotropic state, orientation of LCs is randomized to induce a shrinkage in the direction parallel to the initial LC alignment and expansions in the directions perpendicular to the initial LC alignment.

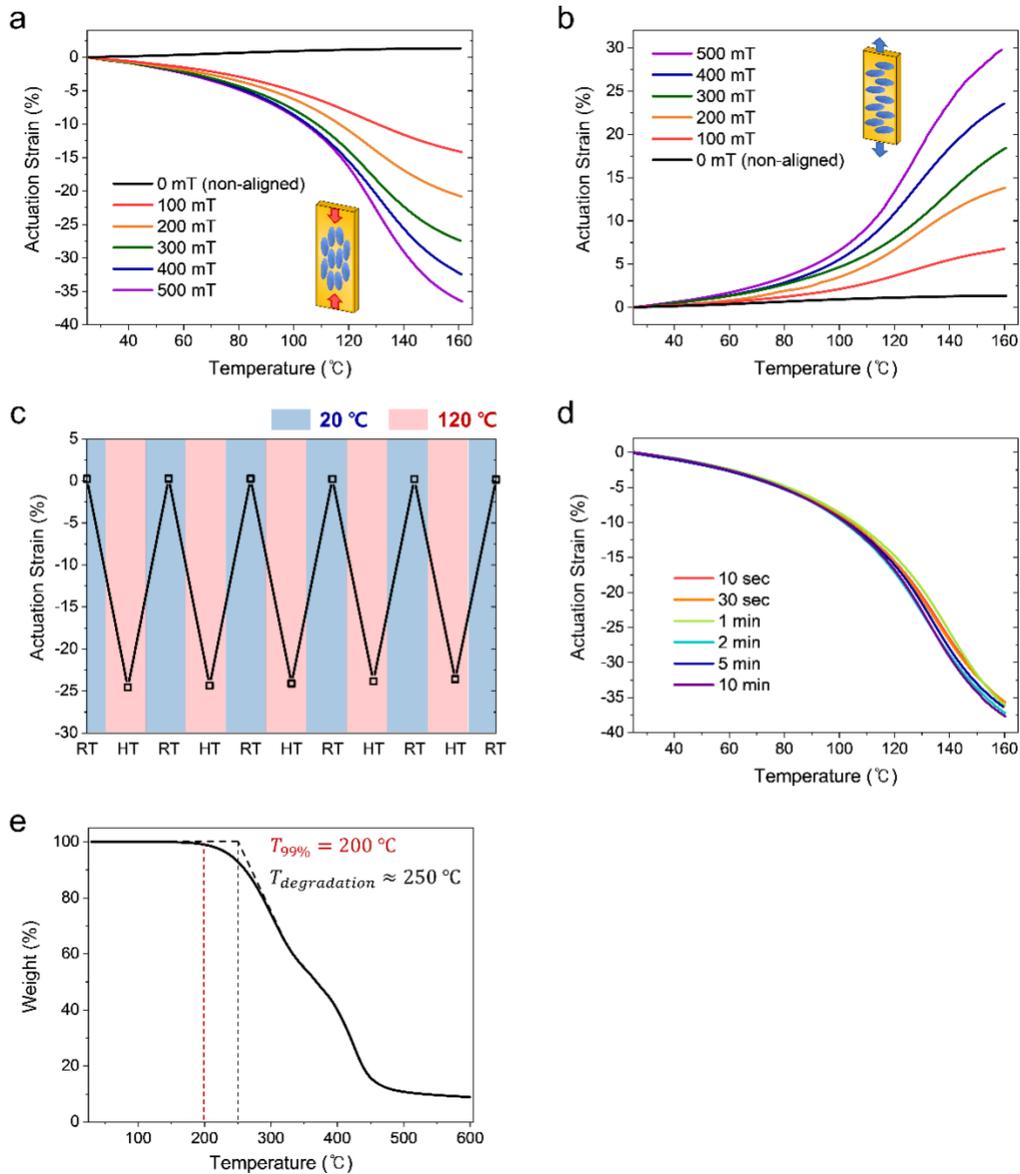

**Fig. 2. Material properties and thermal actuation of LCE samples with magnetic-field alignment. a**, Iso-stress tests of LCE films with parallel alignment under distinct magnetic field strengths, showing a monotonously increasing shrinkage over magnetic field strength. **b**, Iso-stress tests of LCE films with perpendicular alignment under distinct magnetic field strengths, showing a monotonously increasing expansion over magnetic field strength. **c**, Demonstration of repeatable LCE actuation with parallel alignment in temperature cycling. **d**, Iso-stress tests of LCE films with parallel alignment cured after distinct waiting times at 500 mT, indicating an instantaneous and adequate magnetic-field alignment of LCs at room temperature. **e**, TGA result of cured LCE sample, confirming the high thermal stability of the cured LCEs in our studied temperature range.

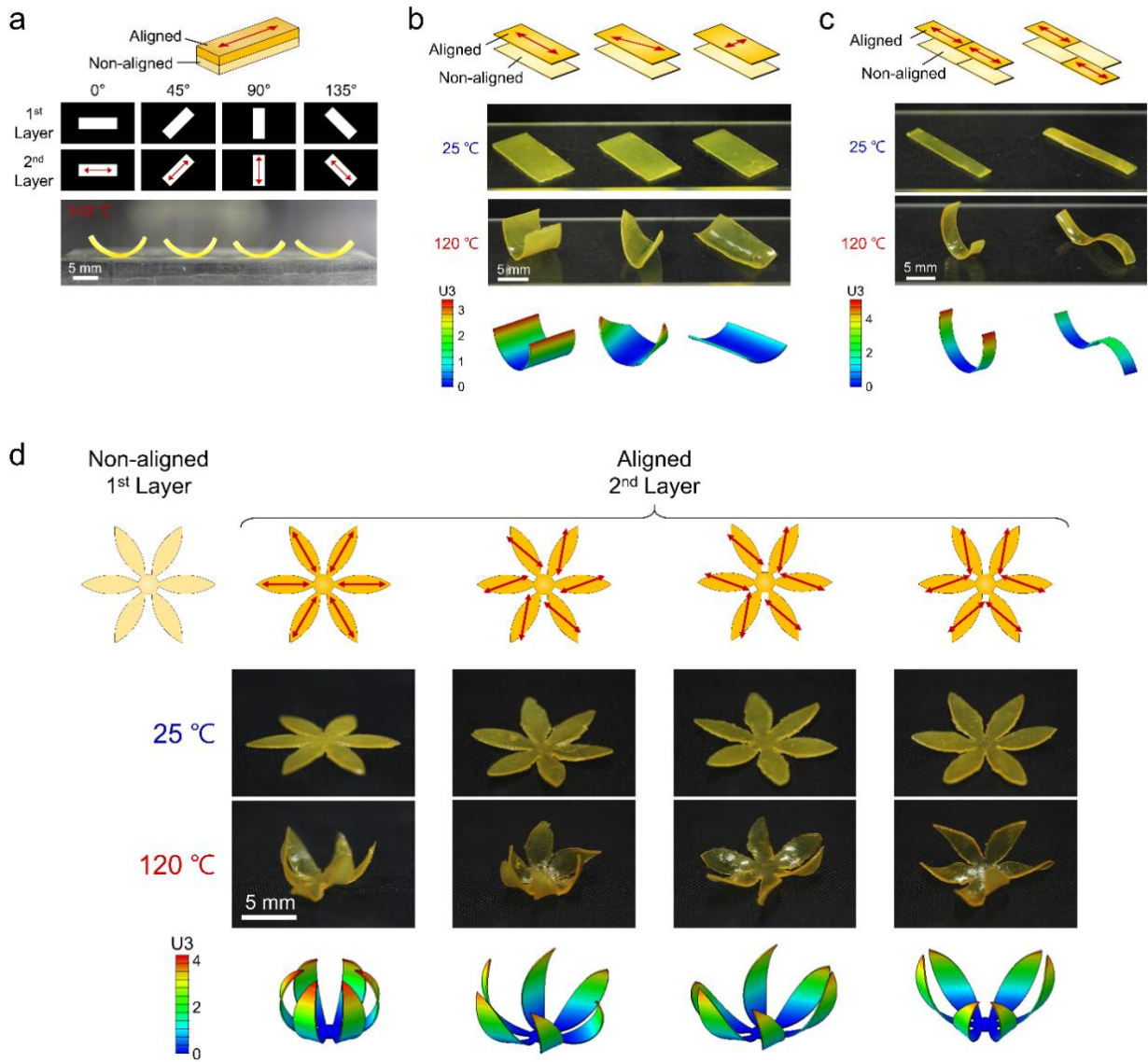

**Fig. 3. Characterization of DLP printed LCE samples with magnetic field molecular orientation programming. a**, Dependence of molecular alignment on magnetic field orientation. After printing the non-aligned bottom layer, the aligned top layer is printed with the magnetic field orientation of 0 °, 45 °, 90 °, and 135 °, respectively. Identical bending behaviors of bilayer films printed at different angles are observed upon heating. **b**, Bending demonstration of LCE films with their top layer aligned by different magnetic field orientation (0 °, 30 °, and 90 ° with respect to the film length direction). **c**, Distinct bending deformation of LCE films with arbitrary LC

alignment locations. **d**, Blooming behaviors of bilayer flower structures with different LC alignments in petals.

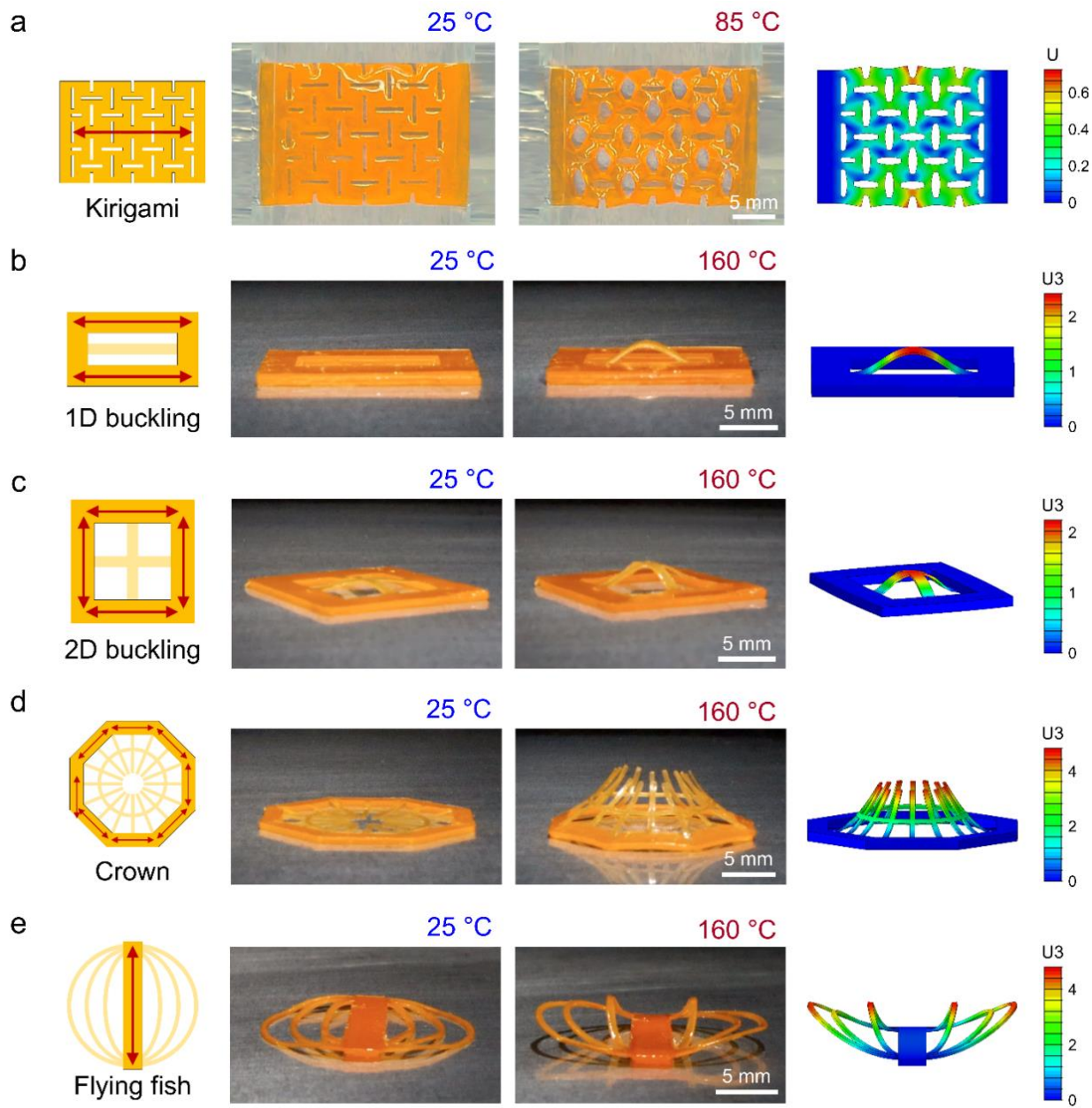

**Fig. 4. Diverse shape morphing LCE structures with programmable LC alignments. a**, Thermal deformation of a LCE kirigami structure. The LCE structure is realized with uniaxially aligned kirigami patterning film. **b**, 2D to 3D deformation of a 1D buckling structure. The LCE structure is realized with the thick uniaxially aligned frame and non-aligned thin central film. **c**, 2D to 3D deformation of a 2D buckling structure. The LCE structure is realized with the biaxially

aligned thick frame and thin central cross non-aligned. **d**, 2D to 3D deformation of a crown structure. The LCE structure is achieved with the aligned frame and non-aligned ring and inner strands. **e**, 2D to 3D deformation of a flying fish structure. The LCE structure is achieved with the aligned central bar and non-aligned thin wings.

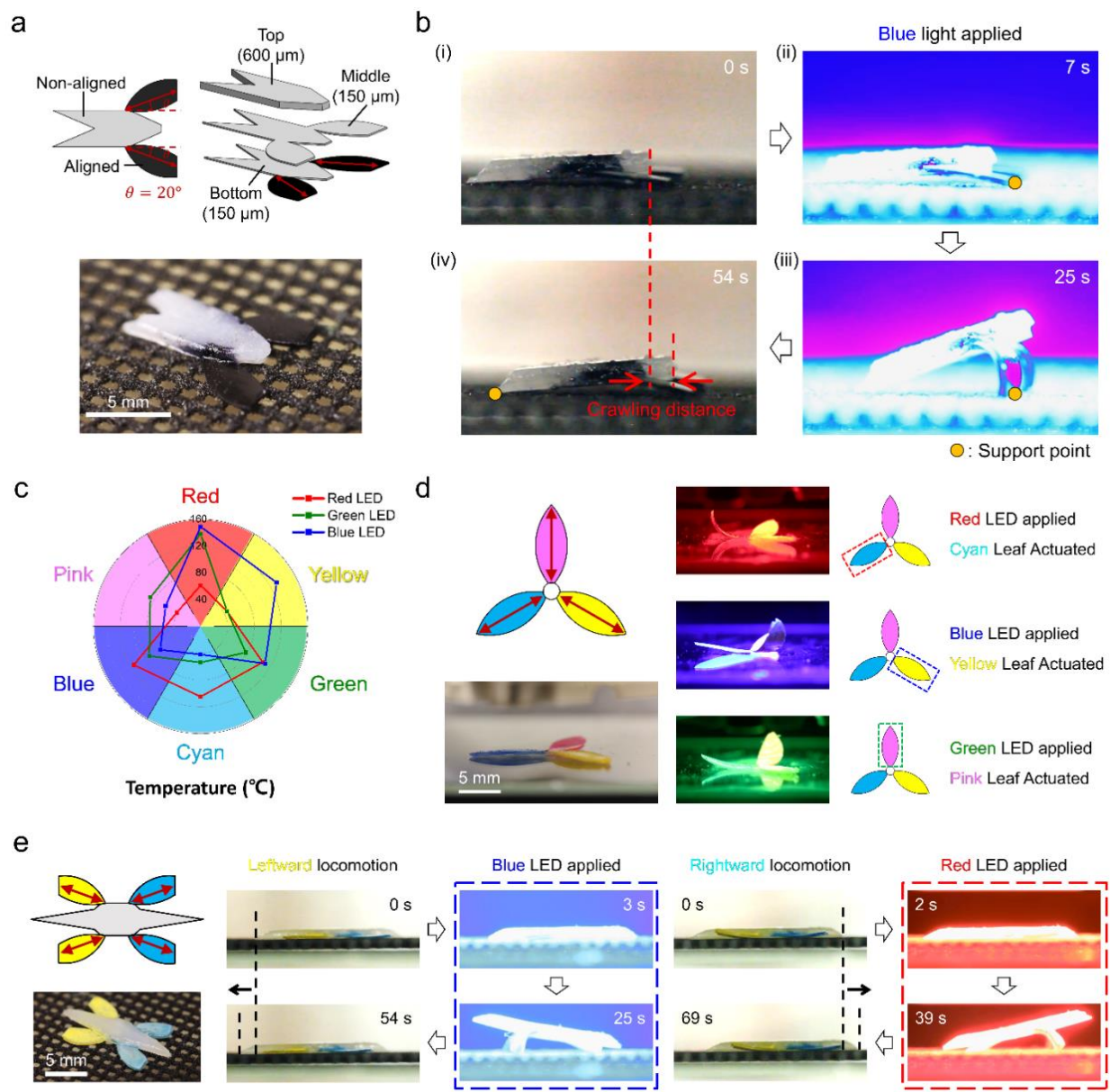

**Fig. 5. Photo-thermal actuation of the LCE structures with programmable LC alignments. a**, Design and actual prototype of the LCE crawler. **b**, Forward crawling locomotion of the crawler under an irradiation. **c**, Radar plot showing the temperature of different colored LCE films under red, green, and blue LED lights. **d**, Demonstration of selective photo-thermal actuation. Exclusively, cyan leaf can be actuated by red light, yellow leaf can be actuated by blue light, and pink leaf can be actuated by green light. **e**, Two-way crawling locomotion of the crawler under red and blue light irradiation, respectively.